**Cometary records revise Eastern Mediterranean chronology around 1240 CE**

Koji Murata,[1,6,*] Kohei Ichikawa,[2,3,*] Yuri I. Fujii,[1,4] Hisashi Hayakawa,[1,5] Yongchao Cheng,[1,6] Yukiko Kawamoto,[6] Hidetoshi Sano[7]

[1] Institute for Advanced Research, Nagoya University, Furo-cho, Chikusa-ku, Nagoya 464-8601, Japan

[2] Frontier Research Institute for Interdisciplinary Sciences, Tohoku University, 6-3 Aramaki, Aoba-ku, Sendai 980-8578, Japan

[3] Astronomical Institute, Graduate School of Science, Tohoku University, 6-3 Aramaki, Aoba-ku, Sendai 980-8578, Japan

[4] Graduate School of Science, Nagoya University, Furo-cho, Chikusa-ku, Nagoya 464-8601, Japan

[5] Institute for Space-Earth Environmental Research, Nagoya University, Furo-cho, Chikusa-ku, Nagoya 464-8601, Japan

[6] Graduate School of Humanities, Nagoya University, Furo-cho, Chikusa-ku, Nagoya 464-8601, Japan

[7] National Astronomical Observatory of Japan, 2-21-1 Osawa, Mitaka, Tokyo 181-8588, Japan

*Email: murata.koji@e.mbox.nagoya-u.ac.jp, k.ichikawa@astr.tohoku.ac.jp

**Abstract**

Eirene Laskarina, empress of John III Batatzes of the exiled Byzantine Empire of Nicaea (1204–1261 CE), was an important Eastern Mediterranean figure in the first half of the thirteenth century. We reassess the date of Eirene's death, which has been variously dated between late 1239 and 1241, with the understanding that narrowing the range in which this event occurred contributes much to understanding the political situation in the area around 1240. George Akropolites, a famous official of the Empire, gives an account that connects Eirene's death to a comet that appeared 'six months earlier', thus pointing to two comet candidates that were visible from the Eastern Mediterranean between 1239 and 1241, one recorded on '3 June 1239' and the other on '31 January 1240'. Recent historians prefer the former, based on historical circumstances and without a critical assessment of the comet records. We revisit the historical records and reveal that the '3 June 1239' candidate was not a comet. On the other hand, the other candidate, sighted on '31 January 1240', was a comet, as supported by multiple historical records in multiple regions, and is also a good fit with Akropolites's narrative. Therefore, we conclude that Eirene died six months after the comet that was seen on 31 January 1240, which places her death in the summer of 1240. Given that the date of her death is crucial for determining some other contemporary events across the Eastern Mediterranean, our results offer a solid basis for further research on the thirteenth-century Eastern Mediterranean.

**1. Introduction**

Astronomical phenomena recorded in various past periods all over the world have sometimes been regarded as keys to determining the dates of major historical events that have been associated with these phenomena in historical writings; such phenomena can also be helpful in evaluating the writings themselves, in terms of their chronological accuracies, the periods in which they were written, the places where they were written, and so on (e.g. Walker 1999; Hayakawa et al. 2017; Hayakawa et al. 2020). In the Byzantine Empire (Late Antiquity until 1453), scholars who inherited ancient Greco-Roman astronomy and astrology maintained the tradition of charting horoscopes and recording notable astronomical phenomena in their writings; they interpreted these phenomena as omens of various future terrestrial events, such as famines, plagues, political crises, and personal life events (Caudano 2020). Modern historians have utilised these records to build a solid chronology of the Empire. For example, a horoscope that was cast for Emperor Constantine VII Porphyrogennetos offers the exact period of his birth, i.e. on the morning of 3 September 905 (Pingree 1973). There are still many such associations between astronomical records and historical events on Earth to be explored. In this paper, we will take one case from the thirteenth-century Eastern Mediterranean.

Eirene Laskarina, the empress of John III Batatzes of the exiled Byzantine Empire of Nicaea that flourished in the Eastern Mediterranean after the Fourth Crusade, which captured the capital of the Empire, Constantinople (1204–1261 CE), was a major figure in that region in the first half of the thirteenth century. As a governor rather than a mere wife, she was an active participant in the Empire's politics, and she contributed to the growth of its power, which led to the recovery of Constantinople in 1261 (Mitsiou 2011). However, the exact year of her death has not been determined. The date of her death is key to defining the death dates of the other two major figures in that period, namely Ivan Asan II, ruler of Bulgaria, and Manuel Angelos, ruler of the Despotate of Epiros (Macrides 2007, pp. 211–214). Moreover, the date of her death is also essential to understanding the inter-political relationships in the Eastern Mediterranean around 1240.

The incontestable time span of her death has been poorly constrained, from June

1239 to May-June 1241. While the former is the date of the partial solar eclipse that she herself is known to have observed on 1239 June 3 (Pingree 1964, p. 367; Stephenson, 1997, pp. 398–399; Macrides 2007, p. 212), the latter is the *terminus ante quem* for the remarriage of her husband, Emperor John III, to Anna of Hohenstaufen (Kiesewetter 1999).[1] The only clue to narrowing down the date of Eirene's death is an account from George Akropolites, who was a famous court official in the thirteenth-century Byzantine Empire. In his *History*, which narrates the events that concerned the Empire from 1204 to 1261, Akropolites relates Eirene's death as follows:

> The empress Eirene died also, a woman both temperate and regal who exhibited imperial majesty greatly. […] When an eclipse occurred [3 June 1239], as the sun was passing through Cancer, around midday—since, when it happened, I had arrived at the imperial residence (the emperor with the empress were residing near a place which they call Periklystra [near Smyrna])—she asked me the reason for the eclipse. […] As I said, this empress died; I think that the eclipse of the sun presaged her death. **A comet also appeared six months earlier in the northern parts. It was a bearded star and lasted three months, appearing not in one place but in several**. (Heisenberg & Wirth 1978, I. pp. 62–64; tr. by Macrides 2007, pp. 210–211 with slight modifications)

According to Akropolites, Eirene died six months after a comet appeared in the northern part of the sky (cf. Tihon 2006, pp. 267–270). In general, modern scholars agree on the accuracy and reliability of *History* (Macrides 2007, pp. 39–41). Thus, it is crucial to identify this comet in order to fix the chronology. In that vein, this comet has been controversially identified with two different astronomical reports that Grumel (1958, p. 474) catalogued as having occurred on '3 June 1239 ([observed in] Europe)' and '31 January 1240 ([observed in] general)'.

    The date of Eirene's death used to be often referenced as in 1241 without taking the comet record into consideration (de Muralt 1871, p. 360; Heisenberg 1903, I. p. 62 and II. p. iv; Heisenberg & Wirth 1978, I. p. 62 and II. p. iv). Later, Brezeanu (1974) discussed the comet record and dated this comet to 31 January 1240 because the comet

---

[1] Angelov (2019, 331–333) argues that the remarriage of John III must have been dated in late August or September 1240, based on the assumption that it was celebrated during the Dog Star was wandering in the sky and in the Virgo season preceding May-June 1241 when the emperor assaulted on Constantinople. The references to the Dog Star and the Virgo are attested in a contemporary work composed by Theodore Laskaris, John's son. In fact, however, the Dog Star seems to refer to the period of the emperor's expedition to Constantinople *after* the wedding, not of the wedding itself. Moreover, the work just says that the Empire's assault was made *before* the Virgo. Thus, these references in Theodore's work only offer another date of John's expedition, July-August 1241, which is another *terminus ante quem* for John's remarriage.

that appeared in June 1239 'could be seen only in Europe' (and not in Asia Minor, where Akropolites would have been located), hence, Eirene's death is estimated to have taken place in the summer of 1240. Madgearu (2017, p. 225) also nominates the January 1240 comet, but his argument is not well-grounded. On the other hand, Macrides (2007) prefers 3 June 1239 for the date of the comet's appearance (and thus she dates her death to the end of 1239) because if the empress died in the summer of 1240 as Brezeanu (1974) claimed, then at the end of 1240/early 1241, it would have been too early for her husband to remarry (see also Munitiz 1984, p. 21).

In recent scholarship, Macrides' dating (2007) has gradually gained consensus without a critical assessment of the records pertaining to the two comets. Furthermore, the most comprehensive catalogue of comets so far does not record Akropolites's comet (Kronk 1999). Therefore, we revisit these two comet candidates from both historical and astronomical perspectives. Contrary to previous scholarship, our investigation shows that the '3 June 1239' candidate seems to not have been a comet at all. In addition, our analysis reveals that the other candidate, '31 January 1240', is a good fit with Akropolites's narrative. Finally, we evaluate the historical implications of the proposed date of Empress Eirene's death.

## 2. Two candidates for the Akropolites comet
### 2.1. Comet Candidate I (3 June 1239)

The 3 June 1239 'comet' that Grumel (1958) listed is based on a record in *Annali della città di Cremona* (*Annals of the City of Cremona*), which was compiled by sixteenth-century Cremonan chronicler Lodovico Cavitelli (here, Grumel (1958) relied on this information as presented by Pingré (1783), p. 403; however, Kronk (1999) does not list it). Using earlier records that have now been lost, in combination with his own experiences, he describes the history of Cremona (45° 8' N, 10° 1' E), from its beginning to 1583. In the entry for the year 1239, he describes the appearance of this star as follows:

> On 3 June [1239], a bearded star together with a celestial object such as a torch was observed, which made haste to the west. They presaged a serious lack of provisions shortly after, and infinite things perished. (Cavitelli 1588, p. 85; tr. by the authors)

However, the following caveats must be noted in the context of identifying this as Eirene's comet. First, this report is found only in a very late compilation, without appearing in any earlier chronicles. Moreover, since, up to the present, no earlier source on which the report was based has been discovered, we are prevented from verifying its reliability (cf. Ditchfield 1995, p. 301). Second, we need to be cautious as to whether this phenomenon was indeed a comet because a relatively rapid westward motion is described. Although, when they are observed over a series of nights, comets exhibit independent trajectories compared to other stars, it is very difficult for the naked eye to recognise movements such as 'from the east to the west' during one night. In addition, the duration of this event does not seem to have been as long as 'three months', as Akropolites observed.

Moreover, Cremona is considerably far away from the Empire of Nicaea, where Akropolites stayed. Although a great comet can be observed simultaneously at distant locations, no such parallel records have been discovered in the case of Candidate I, except for a comet-like astronomical event that was observed in Japan and recorded in the medieval Japanese narrative *Azuma-Kagami*, which describes a 'vapour' that was presumably sighted in Kamakura (35° 19' N, 139° 33' E) during the period 27–30 May 1239, which is a few days prior to Cavitelli's record. The relevant excerpt from *Azuma-Kagami* reads as follows:

> On Day 23 in the 4th month [= 27 May 1239 in the Julian calendar], a fair sky. Ghostly vapour was observed in the northwest during the hour of the dog [= 7 p.m. to 9 p.m.]. The radiance pointed to the southeast, with a length of 8 *shaku* [= ca. 240 cm.], the width was 1 *shaku* [= ca. 30 cm], and the colour was white and red. Although there was no main star, the light shone red as a twitch fire. [...] It disappeared in a while. [...] On Day 26 in the 4th month [= 30 May], a fair sky. It rained from the hour of the tiger [= 3 a.m. to 5 a.m.]. Some ghostly vapour was observed before that. Astronomers held a discussion on the existence of the main star. (Kuroita et al. 1933, pp. 240–241; tr. by the authors)

While Kanda (1935, p. 532) listed this report as a comet sighting, Kronk (1999) did not include it in his comet catalogue. Indeed, it is unclear whether the event was a comet sighting, as 'the main star' was not observed and the phenomenon appeared within only two days. The date of this Japanese report is close to but does not coincide with that of the 3 June 1239 Cremonan star record, hence there is no encouraging factor for regarding these two records as referring to a single great comet.

Finally, in Akropolites's *History*, the description of Eirene's comet was separated from that of the solar eclipse on 3 June 1239, which is the same date as the Cremonan star report; the fact throws doubt on the coincidence of that two phenomena. Overall, unless otherwise supported by an independent report from Asia Minor in that period, it is far from robust to associate this Cremonan star report with Eirene's comet.

Its quick motion, localised appearance, and the plausibility of its short duration are all characteristics that are more consistent with meteors or bolides. Meteors have long tracks, and when they are bright enough, they are often compared with torches and beams (Dall'Olmo 1978; Dall'Olmo 1980, p. 20). Indeed, some meteors outshine the full moon ($m$ = -12.5) (Belton et al. 2004, p. xv) and can be fairly compared with torches. These meteor features are in alignment with the 'star' that was seen in Cremona.

**2.2. Comet Candidate II (31 January 1240)**

Various contemporary writers across the Eurasian continent made records of the comet that was first discovered in the end of January. It is known as C/1240 B1 (Europe: Luard 1869, p. 431; Luard 1876, p. 4; Storm 1888, p. 133; Botteghi 1914–16, p. 15; Garufi 1936–38, p. 205; Stahlman 1952; Fiorese 2004, p. 216. Japan: Kuroita et al. 1933, pp. 252–257; Sasakawa & Yano 1935, pp. 17-37; Kuroita et al. 1929, pp. 186-187; for other minor accounts, see Kanda 1935, pp. 532–536. China: Tuotuo et al. 1977, ch. 42 and 56. For Japanese and Chinese accounts, see also Pankenier et al. 2008, pp. 147–149 and 563). This comet was reportedly visible in the northern hemisphere from 27 January to 31 March 1240 (see in general Kronk 1999, pp. 215–217).

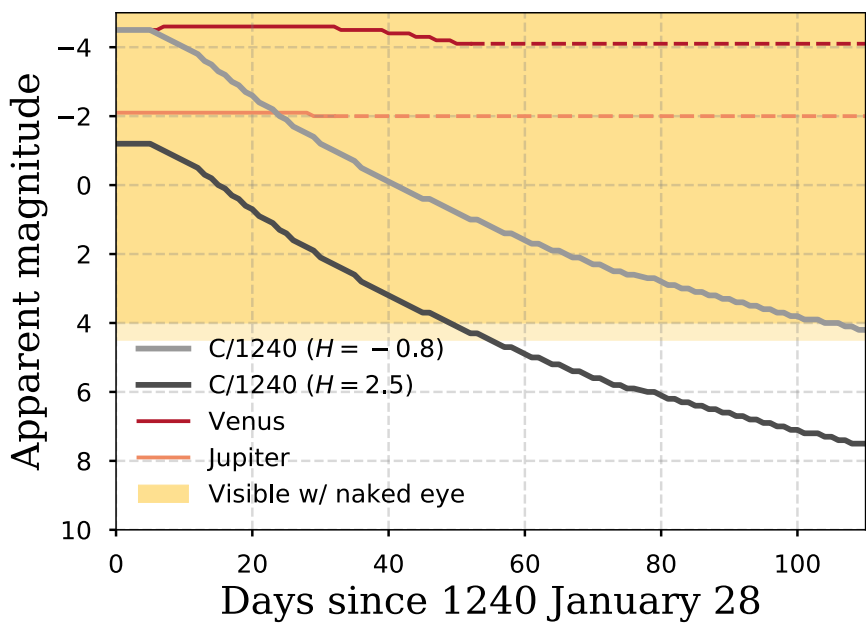

Figure 1. The theoretical light curves of C/1240 B1 in 1240. The black and gray solid line respectively represents the theoretical light curve of C/1240 B1 assuming the fiducial absolute magnitude of H=2.5, and Venus matched magnitude of H=-0.8, which is based on the description of *Azuma-Kagami*. The apparent magnitudes of Jupiter (m~-2.1 to -2.0) and Venus (m~-4.5 to ~-4.1, which depends on the observing dates)

are also shown with orange and red solid-line, connected with the dashed-line where Jupiter/Venus is not visible in the night. The yellow area represents the visible area with the naked eye (limiting magnitude of m=4~4.5).

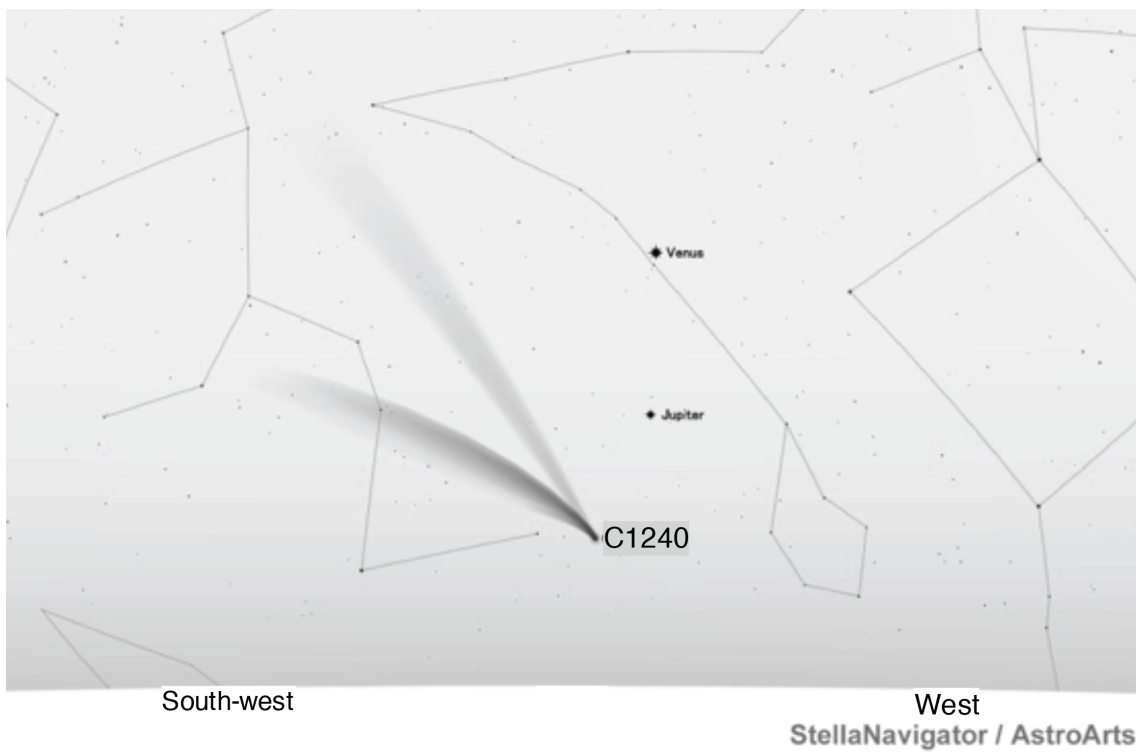

Figure 2. The sky on January 31, 1240 at 7 p.m. (LMT) at 38°26' 0" N, 27°9' 0" E, as produced by Stellar Navigator 11/AstroArts.[2]

When we reproduced the appearance of the comet C/1240 B1 from Smyrna (38° 26' N, 27° 9' E), which was Akropolites's supposed observational site, we adopted the orbital elements that appear in Marsden and Williams (2008), as calculated by Ogura (1917). In late January 1240, the comet C/1240 B1 was probably not visible until around January 27 as the sky was too bright due to sunlight. When it did become visible, it would have been seen slightly above the horizon for less than approximately one hour in the twilight before it sank into the ocean to the west of the city. When adopting H=2.5 as the comet's absolute magnitude (Kronk 1999), C/1240 B1's apparent magnitude around the time is -1.2 (Figure 1), which is a little darker than Jupiter ($m$ = -2.1), which should also appear in the western sky. Figure 2 is a representation of the sky on January 31, 1240 at 7 p.m. (LMT). Over time, the comet remains in the sky for a longer duration, but it becomes progressively darker, and the magnitude drops to 0 on February 12. Its location shifts toward north as described by Akropolites (Figure 3). By virtue of its tail, it was still recognisable at the end of March, although the magnitude drops to 5. It is questionable whether this comet was visible for about three months because the coma's brightness is as dark as $m = 6.7$ three months after the comet first became detectable at the observational site. However, it could have still been recognisable, albeit barely, depending on conditions such as the composition, porosity, and size of the comet's nucleus as well as solar activity, and so on.

---

[2] https://www.astroarts.co.jp/products/stlnav11/index-j.shtml

In *Azuma-Kagami*, there is a description that the coma was as big as Venus on February 1 (Kuroita et al. 1933, p. 253; Pankenier et al. 2008, p. 147), which is much brighter than the theoretical lightcurve of H=2.5 (Figure 1). This implies the possibility that the comet was actually brighter than the expectation based on Kronk (1999) or it was due to cometary/solar activity. Figure 1 also shows the theoretical lightcurve assuming H=-0.8, which matches the magnitude with Venus on February 1. In this case, the comet was visible with the naked eye for three to four months.

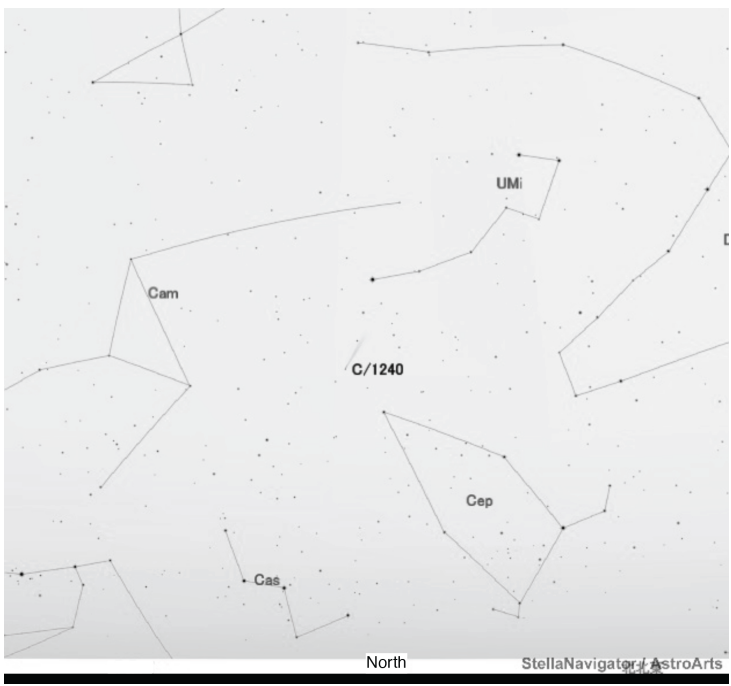

Figure 3. The appearance of C/1240 B1 on March 31 at 10 p.m. (LMT), as represented by Stellar Navigator 11/AstroArts. The magnitude of the comet is 5.1, and stars with $m \leqq 5.5$ are displayed. The point of observation is the same as in Figure 2.

Based on the above results, we conclude that Comet Candidate I was actually a meteor or a shooting star event, and therefore the Candidate I event is not the one that Akropolites described. Meanwhile, Comet Candidate II was a great comet event that lasted a few months and has features that are consistent with those Akropolites described. Finally, it must be noted that Pingré (1783, pp. 403–404) already suggested that Akropolites's comet was the comet of 1240, although his reference to it has not been consulted thus far.

## 3. Discussion
### 3.1. Is there a possibility that bright, visible comets appear twice a year?
One might argue that we may have overlooked other comets rather than C/1240 B1 (Candidate II in Section 2.2.). Hence, it may be worth reviewing how often visible comets can be observed within a year. Here, we assume that there was another great comet that was visible for a few months in the same year. Based on historical records of the great comets that are visible to the naked eye for a few months (e.g. Kronk 1999), the frequency of such great comets is roughly once per decade somewhere on the Earth. If we pinpoint one place, the probability reduces by a factor of 2x2=4 for two reasons: because we can see only half of the sky from one place and due to the limited season at a particular observation time. Therefore, one great comet can be observed every 40 years. Hence, the likelihood of observing two great comets in a single year would become $P(X = 2) \sim e^{-1/40} * (1/40)^2 / 2! \sim 3 \times 10^{-4}$, which represents a negligibly small chance.

In addition, if such a great comet had swung-by the Sun, historical records of the event should exist, considering that there are several historical records of C/1240. Therefore, we conclude that it is unlikely that the phenomenon Akropolites mentioned was actually a different comet that appeared within the same year, and thus C/1240 would be the unique astronomical event that corresponds to his narrative.

### 3.2. The historical implications
Our examination reveals that Akropolites's comet is more consistent with C/1240 B1, which was first observed in late January 1240. On the other hand, as a comet-sighting date, June 1239 contradicts his description. Therefore, Empress Eirene most likely died during the summer of 1240, six months after the appearance of the comet at the end of January 1240. This date is important to (re)considering several political affairs across the Empire of Nicaea around 1240.

First, Eirene's death date approximates the death dates of two neighbouring rulers, to which Akropolites's *History* also provides a chronological clue. One is Manuel Angelos, the ruler of the Despotate of Epiros, who died after 'not much time had passed' following his reconciliation with his brothers in 1238/39 (Heisenberg & Wirth 1978, p.

62; Macrides 2007, p. 210; Stefec 2015, p. 50). Since Manuel's death is narrated in the context of Eirene's death, their death dates have been placed close together at around 1240. Another case is that of Ivan Asan II, who ruled Bulgaria beginning in 1218. According to Akropolites, he died 'a short time' after Eirene (Heisenberg & Wirth 1978, p. 64; Macrides 2007, p. 211). Our finding places the date of Asan's death between the summer of 1240 and 24 June 1241, when, as his father's successor, Asan's son Kaliman reached a truce with John III Batatzes and Baldwin II of Constantinople (Dölger & Wirth 1977, p. 36; cf. Madgearu 2017, pp. 225–226).

   Second, the empress's death date leads us to reconsider the meaning of the remarriage of her husband, Byzantine Emperor John III Batatzes, to Constanza-Anna, a daughter of Frederic II of Hohenstaufen, in late 1240/early 1241 (Kiesewetter 1999). Given that the emperor remarried only a few months after Eirene's death, there should be discernible reasons for his haste. John III may have hastened his remarriage in order to strengthen a desirable alliance with Frederick II of Hohenstaufen. One of the most probable motives for this was the Mongols' advancement towards Eastern Europe and Anatolia around 1240 (Korobeinikov 2014; Murata 2015). The other was the papacy's hostile attitude towards both Frederick II and John III, as evidenced by the excommunication of both rulers and pending crusades against them, especially after Pope Gregory IX excommunicated Frederick II for the second time in early 1239 (Murata in press). Curiously enough, no contemporary Byzantines recorded the emperor's remarriage, despite its importance to the political situation across the Mediterranean world in that period. Indeed, in his *History,* George Akropolites is silent on the matter of John III's remarriage, and he mentions Constanza-Anna by name only once, in the context of John III's adulterous relationship with one of her ladies Marchesina (Heisenberg & Wirth 1978, p. 104; cf. Macrides 2007, pp. 49, 56, 271, 275–6). Another influential contemporary in the Byzantine political sphere, Nikephoros Blemmydes, anonymises Constanza-Anna in his writings, while explicitly criticising the lady Marchesina (Munitiz 1984; Munitiz 1988). Modern scholars have postulated that these silences and antipathies with respect to the emperor's remarriage were due to John III's adulterous affair with Marchesina. However, we can now suppose that the strikingly short mourning period for the deceased Empress Eirene contributed to the emperor's subjects' growing distrust of him. Contrary to the reception of the new empress, Constanza-Anna, Eirene's subjects honoured and grieved for her (Bury 1901; Hörandner 1972; cf. Macrides 2007, pp. 20–21), and her achievements were remembered by later Byzantine historians, including Nikephoros Gregoras (Schopen 1829, pp. 44–45; van Dieten 1973, pp. 41–42, 85–86).

## 4. Conclusion

We have revisited the two comet records of '3 June 1239' (Comet Candidate I) and '31 January 1240' (Comet Candidate II), motivated by the need to revise the date of death of Empress Eirene Laskarina, wife of John III Batatzes, which has been poorly constrained between late 1239 and 1241. Having examined the relevant historical records, we arrived at the following key findings:

1. The historical record of the '3 June 1239' event revealed that it was a very short astronomical event that lasted one night, with the phenomenon moving towards the 'east' in an independent direction, compared with the trajectory of other stars. This suggests that the '3 June 1239' event was not actually a comet but was more likely either a meteor or a bolide, thus necessitating a re-consideration of Eirene's death date.
2. Since it has been determined that Comet Candidate II, identified as C/1240 B1, should have been visible ($m < 5$) for a few months, this comet is largely consistent with George Akropolites's account of his observed phenomenon, which he describes as 'lasting three months', thus placing Eirene's date of death in the summer of 1240 and not the winter of 1239/1240.
3. Eirene's revised date of death provides a strong basis for constraining the death dates of two contemporary rulers in the Balkans and for reconsidering the political affairs of her husband, John III Batatzes, around 1240.

**Acknowledgements**

We thank Atsushi Naruko, Shunsuke Kosaka, and Kohji Tsumura for several fruitful discussions. This work was supported by the Programme for Establishing a Consortium for the Development of Human Resources in Science and Technology, Japan Science and Technology Agency (JST); the 2019 YLC Collaborate Research Grant, Nagoya University; and the Japan Society for the Promotion of Science (JSPS) KAKENHI (K. Murata: 19K13389; K. Ichikawa: 18K13584, 20H01939).